\documentclass{rstransa}

\begin{document}
\title[Irreversibility and chaos]{A Semiclassical Reversibility Paradox in Simple Chaotic Systems}
\author[]{Steven Tomsovic}
\address{Max-Planck-Institut f\"ur Physik komplexer Systeme, N\"othnitzer Stra\ss{}e 38, D-01187 Dresden, Germany, and Department of Physics and Astronomy, Washington State University, Pullman, WA USA 99164-2814}
\subject{reversibility paradox}
\keywords{reversibility, chaos, quantum, classical, semiclassical}

\begin{abstract}
Using semiclassical methods, it is possible to construct very accurate approximations in the short wavelength limit of quantum dynamics that rely exclusively on classical dynamical input.  For systems whose classical realization is strongly chaotic, there is an exceedingly short logarithmic Ehrenfest time scale, beyond which the quantum and classical dynamics of a system necessarily diverge, and yet the semiclassical construction remains valid far beyond that time.  This fact leads to a paradox if one ponders the reversibility and predictability properties of quantum and classical mechanics.  They behave very differently relative to each other, with classical dynamics being essentially irreversible/unpredictable, whereas quantum dynamics is reversible/stable.  This begs the question, ``how can an accurate approximation to a reversible/stable dynamics be constructed from an irreversible/unpredictable one?''  The resolution of this incongruity depends on a couple key ingredients, a well-known, inherent, one-way structural stability of chaotic systems and an overlap integral not being amenable to the  saddle point method.

\end{abstract}

\maketitle

\section{Introduction}

In the latter period of the 1800's, Loschmidt's objection~\cite{Loschmidt76} to Boltzmann's work on the statistical mechanical foundations of the second law of thermodynamics~\cite{Boltzmann72} gave rise to a reversibility paradox in which it appears that it should not be possible to use a microscopically reversible dynamics to infer thermodynamic irreversibility, yet both were well established.  A broader subject of reversibility has developed, including but not limited to establishing various so-called `arrows of time'~\cite{Eddington28}, that continues to be of great interest today in many realms~\cite{Sokolov08,Taddese10,Garciamata11}.  The subject of interest in this contribution, however, is in regard to a very different aspect of reversibility than the Loschmidt paradox, that amusingly involves the opposite direction -- irreversible to reversible, and is not concerned with the thermodynamic limit, but rather the correspondence between quantum and classical dynamics of strongly chaotic systems.  It turns out that the Correspondence Principle~\cite{Bohr20} is deeper and more subtle than the view that unfolds strictly from the assertion that classical dynamics must emerge from quantum dynamics in some limit, i.e.~ $\hbar \rightarrow 0$ or `just replace commutators with Poisson brackets.'  In fact, the reverse connection exists, i.e.~semiclassical theory uses purely classical dynamical information to construct more and more accurate approximations of quantum dynamics in this very same limit~\cite{Maslov81}.  

The reversibility paradox of interest here arises because the properties of classical and quantum dynamical systems behave so differently, yet are so deeply connected through this semiclassical construction.  We have in mind simple systems which are strongly chaotic, bounded, and for which there are no environmental or other degrees of freedom.  We also consider only evolution under Hamilton's or Schr\"odinger's equations and do not introduce measurement.  Such systems' quantum realizations possess unitary dynamics which are inherently reversible.  After all, the inverse dynamics are given by merely taking the Hermitian conjugate of the unitary propagators that describe the forward dynamics.  In addition, the forward dynamics is stable since knowing the unitary propagator for some time interval leads only to matrix multiplication to obtain the dynamics for any other multiple of the interval.  In contrast, the classical dynamical picture is not so straightforward.  Although, the dynamics are predictable and reversible with an exact representation of the state of the system and an exact implementation with a perfectly precise set of equations of motion, any deviation of either leads to exponential instability in the predictability or the reversed dynamics.   There are  extremely short time scales with logarithmic behavior over which the dynamics is predictable and reversing the system leads to the forward dynamics in retrograde.  This distinction between the quantum and classical dynamics was nicely raised by Shepelyansky and illustrated with a strongly kicked rotor~\cite{Shepelyansky83}, one of the most important paradigms of simple chaotic systems~\cite{Chirikov79}.

A quantity of great interest for this contribution was introduced by Peres~\cite{Peres84}, and which may be called the Loschmidt echo or fidelity depending on author or context.  Interestingly, Peres was unhappy with the circumstance that thermodynamical irreversibility had not been placed on analogous or equivalent conceptual foundations for both quantum and classical dynamics.  He found that by introducing the notion that the Hamiltonian is not perfectly known or controlled, as opposed to the state of the system, he could rectify this situation.  He suggested concentrating on the evolution properties generated by two slightly different systems and comparing how the same initial state's evolution diverges between the two; i.e.~the sensitivity of the system to perturbations is critical. In spite of his motivations, this idea may be invoked to study predictability and reversibility in a much broader context, including those contexts that have nothing to do with the thermodynamic limit.

A variety of logarithmically short time scales emerge in studies of the classical and quantum dynamics of simple chaotic systems~\cite{Shepelyansky83,Berman78,Berry79b,Berry79,Jalabert01,Cerruti02}, depending on the specific purpose.  The classical time scales of greatest interest in this paper are a function of the dynamical entropy~\cite{Kolmogorov58,Kolmogorov59,Sinai59} or maximal Lyapunov exponent~\cite{Lyapunov92}, and perturbation strength.  First, there is a time scale associated with the classical mixing property in which a localized region of phase space on some scale under propagation spreads into all of the available space to the same localized scale as it began.  Secondly, there is the scale over which a classically chaotic dynamics is faithfully reversible either in a computation or using slightly different governing equations (Hamiltonians).  Here the term irreversibility is used to specify that classically chaotic dynamics is reversible in practice only for this incredibly short time scale; it does not imply that there is no exact microreversibiity of the dynamics.  An exponentially increasing amount of information specifying the system and its current state are required to follow its dynamics linearly further forward in time.  

The quantum time scale of greatest interest is the Ehrenfest time and it marks the latest time in the quantum dynamics beyond which interference must begin to appear independently of how well an initial state starts out localized, i.e.~such as a minimum uncertainty wave packet.  The details of the quantum and classical dynamics must cease to correspond beyond this time.  Nevertheless, it has been shown that the semiclassical construction of quantum dynamics for a chaotic system works extremely well, far past the Ehrenfest time, and the long time breakdown of this correspondence is not bounded by the predictability or irreversibility time scales of the classical dynamics~\cite{Tomsovic91b,Oconnor91,Oconnor92,Sepulveda92,Tomsovic93,Tomsovic93b,Heller93s}.  Rather, the construction works accurately for algebraic time scales.  This is far beyond a time regime where the classical dynamics is predictable/reversible, and yet the quantum dynamics is perfectly stable/reversible.  Bear in mind that the distinction between the classical reversibility time scale and the semiclassical validity time scale grows without bound in the $\hbar\rightarrow 0$ limit.  It would almost seem as though one should investigate using the quantum dynamics to improve predictability and reversing the classical dynamics, thereby overcoming both problems.  Of course, this is not possible, but contemplation of this paradox reveals that its resolutions relies on a well-known, inherent, structural stability of unique time direction for the stable and unstable manifolds found in chaotic systems~\cite{Ott02,Cerruti02} and a quantum overlap integral is not amenable to a saddle point approximation (which is the foundational approximation of the semiclassical construction to begin with).

\section{Predictability and Reversibility}

Our attention is focussed on $d$ degree-of-freedom systems that are strongly and fully chaotic, meaning they possess significant positive Lyapunov exponents~\cite{Lyapunov92}, there are no hidden dynamical symmetries, the phase space is overwhelmingly filled with chaotic trajectories, and the system has a finite phase space volume; the possible existence of an extremely small relative volume of stable trajectories does no harm to any of the main ideas.  The system is assumed to exist in both its classical and quantum realms and $d$ is not large enough to be on the thermodynamical scale.  In this section, we review just how different some of the basic classical, quantum, and semiclassical dynamical properties are.

\subsection{Classically chaotic systems}

To set the key time scales for predicability, reversibility, and mixing, a few definitions are indispensable.  Let the Hamiltonian be uncertain by a perturbation
\begin{equation}
H_\epsilon({\bf x}, {\bf p}) = H_0({\bf x}, {\bf p}) + \epsilon H_1({\bf x}, {\bf p})
\end{equation}
where $H_0$ is fixed and generates a strongly chaotic dynamics, and $\epsilon H_1({\bf x}, {\bf p})$ represents the uncertainty in the representation of the system either because of a lack of perfect knowledge or through algorithmic (computational) limitations.   Comparing any trajectory of identical initial conditions propagated both by $H_0({\bf x}, {\bf p})$ and $H_\epsilon({\bf x}, {\bf p})$ leads to a divergence between the two 
\begin{equation}
||({\bf x}(t), {\bf p}(t))_\epsilon - ({\bf x}(t), {\bf p}(t))_0 || \sim \epsilon\ {\rm e}^{\mu_{max} t}
\end{equation}
where $\mu_{max}$ is the maximal positive Lyapunov exponent of $H_0({\bf x}, {\bf p})$.  These two trajectories thus follow each other only over a time scale given by
\begin{equation}
\label{reversal}
\tau_p \sim \frac{1}{\mu_{max}} \ln \frac{1}{\epsilon}
\end{equation}
where the subscript $p$ indicates that this can be thought of as a predictability time scale.  This is not the best analogy to what happens in quantum dynamics.  It would be good to consider a density of trajectories analogous to an initial minimum uncertainty state.  A good option is
\begin{equation}
\rho({\bf x},{\bf p}) = \frac{1}{\left( \pi \sigma_x \sigma_p \right)^{d/2}} \exp \left[ -\frac{({\bf x} - {\bf x_0})^2}{2\sigma^2_x} - \frac{({\bf p} - {\bf p_0})^2}{2\sigma^2_p} \right]
\label{gaussian}
\end{equation}
where the uncertainties, $\sigma_x,\sigma_p$ are tending towards the very small limit, and the normalization is chosen such that $\langle \rho({\bf x},{\bf p}) , \rho({\bf x},{\bf p})\rangle = \int {\rm d}{\bf x} {\rm d}{\bf p} \rho({\bf x},{\bf p}) \rho({\bf x},{\bf p})=1$.  The most evident way of testing predictability is to propagate the above distribution with both $H_\epsilon$ and $H_0$ and calculate the overlap.  In so doing, one would be calculating the classical version of Peres fidelity~\cite{Peres84}, $\langle \rho_\epsilon({\bf x},{\bf p};t) , \rho_0({\bf x},{\bf p};t)\rangle$.  Notice that, in fact, this is equivalent to propagating the density forward in time with $H_0$, reversing the prorogation with $H_\epsilon$, and overlapping that with the initial density.  Thus for our purposes, the reversibility and predictability time scales are identical and captured by the same measure, and they are intimately connected with the system's sensitivity to perturbations.  

Our interest is in the initial decay as opposed to the long time asymptotic behavior, and both have been studied~\cite{Benenti03}.  A heuristic method of calculating the initial classical fidelity decay would be to rely on the normal coordinate form~\cite{OzorioBook} of the central trajectory of $\rho_0({\bf x},{\bf p};t)$.  There are three effects to consider in the forward propagation of $\rho_\epsilon({\bf x},{\bf p};t)$ in those coordinates; a shift captured by the divergence of the central trajectories, a linear transformation of the unstable and stable manifolds, and a change in the Lyapunov exponents.  The latter is negligible compared with the others.  The linear transformation of the manifolds would lead to half the predictability time scale above if it were either a shear or rotation, but this is inhibited by structural stability of the manifolds (more ahead).  However, the effect of a shift for the perturbed system in which the new unstable manifold were ``parallel'' to the unperturbed one does have a marked effect.  Incorporating just this relative shift of the density in the Gaussian integrals gives for the most significant term (good for up to intermediate times),
\begin{equation}
\label{classpredict}
\langle\rho_\epsilon({\bf x},{\bf p};t) , \rho_0({\bf x},{\bf p};t) \rangle \sim \exp\left[ - \alpha^2\epsilon^2\left({\rm e}^{\mu_{max} t}-1\right)^2 \right] \quad ({\rm thus}\ \tau_r = \tau_p \sim \frac{1}{\mu_{max}} \ln \frac{1}{\alpha \epsilon})
\end{equation}
where the parameter combination $\alpha\epsilon$ is the ratio of the shift to twice the initial standard deviation along the stable direction associated with the maximal Lyapunov exponent.  The predicability and reversibility time scales remain unchanged with the introduction of trajectory densities, other than the slight effect of $\alpha$.

Another critical time scale derives from the property of mixing.  Heuristically speaking, the system has an available phase space volume $V(E)$ given by
\begin{equation}
V(E) = \Delta E \int {\rm d}{\bf x} {\rm d}{\bf p}\ \delta\left[ E - H({\bf x}, {\bf p}) \right]
\end{equation}
where $\Delta E$ is the energy effectively spanned by initial conditions contained in the Gaussian density of phase points.  This volume can be divided into a number of localized cells all having the same volume as the Gaussian density of Eq.~(\ref{gaussian}), which is to a good  approximation $\left(4\pi\sigma_x\sigma_p \right)^d$ (volume inside the $2\sigma$ contour).   As $\rho({\bf x},{\bf p})$ is propagated it stretches into a kind of $d$-dimensional hyperplanar object and folds until it gets so complicated that it simultaneously has intersected all the cells.  An important property of strongly chaotic classical dynamics is the rate of dynamical entropy production (named the Kolmogorov-Sinai or KS-entropy $h_{KS}$) as introduced and developed by Kolmogorov~\cite{Kolmogorov58,Kolmogorov59} and Sinai~\cite{Sinai59}, respectively.  It can also be thought of as a measure of information production.  This gets connected to the mixing time scale, which gives the time at which each phase space cell has typically a single intersection with $\rho({\bf x},{\bf p};t)$.  In other words, at the mixing time the number of cells equals the expansion magnitude of the initial cell (the dynamical entropy)
\begin{equation}
N(E) = \frac{V(E)}{\left(4\pi\sigma_x\sigma_p \right)^d} = \exp \left( h_{KS}\tau_m \right)
\end{equation}
Defining a cell number and geometric mean phase space volume per degree of freedom, i.e.~${\cal V}(E) = V(E)^{1/d}, {\cal N}(E) = N(E)^{1/d}$, gives the mixing time scale
\begin{equation}
\label{mixing}
\tau_m = \frac{d}{h_{KS}}\ln \frac{{\cal V}(E)}{4\pi\sigma_x\sigma_p} = \frac{d}{h_{KS}}\ln {\cal N}(E) = \frac{1}{\langle \mu \rangle}\ln {\cal N}(E)
\end{equation}
The interpretation is that the mixing time scale is the inverse of the dynamical entropy production per degree of freedom multiplied by the natural logarithm of the effective number of phase space cells per degree of freedom.  In the last form, it is noted that the dynamical entropy production per degree of freedom is just the mean of the positive Lyapunov exponents for the system due to Pesin's theorem~\cite{Pesin77}.  The structure of the mixing time is quite analogous to the predictability/reversibility time scale with $1/\epsilon$ replaced by ${\cal N}(E)$ and the maximal positive Lyapunov exponent replaced by the mean of the positive exponents ($\langle \mu \rangle =h_{KS}/d$).  As a final comment, one measure of unpredictability in the algorithmic complexity sense leads to the Alekseev-Brudno theorem~\cite{Brudno78,Alekseev81}, which proves the equivalence of the information associated with a trajectory segment asymptotically to $h_{KS}$.  Here, in contrast the predictability time scale definition led to the maximal Lyapunov exponent whereas the mixing time scale involved $h_{KS}$ per degree of freedom.

\subsection{Quantum dynamics of chaotic systems}

There are multiple ways in which authors have defined a quantum mechanical entropy.  Two of the most discussed are by Connes, Narnhofer, and Thirring~\cite{Connes87} and another due to Alicki, Fannes~\cite{Alicki94}, and Lindblad~\cite{Lindblad88}.  As the quantum dynamics under discussion here are strictly unitary, there is no coupling to the environment or external degrees of freedom, the systems are bounded, and measurement is not being considered, there is little need for these entropy definitions, and we can consider the dynamical entropy as vanishing.  The dynamics are reversible and no information is produced.  For studying the time scale of predictability or reversibility, it is sufficient to consider the unitary dynamics.  

Making use of quantum localized wave packets in analogy with Eq.~(\ref{gaussian}) gives
\begin{equation}
\Phi({\bf x}) = \langle {\bf x} | \Phi \rangle = \frac{1}{\left( 2\pi \sigma^2_x \right)^{d/4}} \exp \left[ -\frac{({\bf x} - {\bf x_0})^2}{4\sigma^2_x} +\frac{i}{\hbar} {\bf p_0}\left( {\bf x} - {\bf x_0} \right) \right] 
\end{equation}
Choosing $\sigma_x$ identically for the quantum and classical cases, and choosing $\sigma_p = \hbar/(2\sigma_x)$ matches the full argument of the exponential in the classical density to that of the Wigner transform of this wave packet and satisfies the minimum uncertainty relation.  The classical density volume equates to the Planck cell volume taken up by a single quantum state, $(4\pi \sigma_x \sigma_p)^d = h^d$.  

The time scale for quantum reversibility and predictability has been specified rather precisely in a number of fidelity studies~\cite{Jalabert01,Jacquod01b, Cerruti02,Cerruti03} and the scaling depends on the magnitude of the perturbation.  To present these results most simply, it is convenient to define a dimensionless constant $\gamma$~\cite{Cerruti03},
\begin{equation}
\label{gamma}
\gamma^2 = \frac{\hbar^2}{2g\epsilon^2\tau_H K(E)}
\end{equation}
where the Heisenberg time scale is $\tau_H = h / D(E)= 2 \pi V(E)/(\hbar^{d-1}\Delta E)$ [$D(E)$ being the mean level spacing], $g$ is a discrete symmetry index, and an important quantity, the classical action diffusion constant~\cite{Bohigas95} is
\begin{equation}
\label{diffuse}
K(E) = \int_0^\infty \left\langle H_1[{\bf x}(0), {\bf p}(0)] H_1[{\bf x}(t), {\bf p}(t)]  \right\rangle_{po} dt
\end{equation}
These relations assume time reversal invariance; the generalization is straightforward (divide $g$ by two if noninvariant).  The classical action diffusion constant contains the needed information about how a perturbation alters chaotic orbits' classical actions.  For an arbitrary perturbation, as a trajectory winds through the available phase space it sometimes adds a bit or subtracts a bit of action.  The diffusion constant characterizes this process.  Its definition follows from first order classical perturbation theory.  With these definitions, the time scale for quantum reversibility and predictability can be stated as
\begin{equation}
\tau_r=\tau_p= \left\{\begin{array}{lll}
= \gamma\ \tau_H & \gamma^2 > 1 & {(\rm quantum\ perturbative\ regime)}\\
= g \gamma^2 \tau_H & 1 > \gamma^2 > \frac{d}{h_{KS}\cdot \tau_H} & {(\rm Fermi\ Golden\ Rule\ regime)}\\
= \frac{1}{<\mu>} & \frac{d}{h_{KS}\cdot \tau_H} > \gamma^2  & {(\rm Lyapunov\ regime)}
\end{array}\right.
\label{tscales}
\end{equation}
For the quantum perturbative regime, Peres~\cite{Peres84} gave the result in terms of quantum level velocities in his original paper, but the form quoted here gives all the dependences in terms of purely classical quantities and Planck's constant through the use of semiclassical theory~\cite{Cerruti02, Cerruti03}.  Note that the result given above for the Lyapunov regime~\cite{Jalabert01} would recover the form of Eq.~(\ref{classpredict}), i.e.~$\tau_r \sim -\ln \epsilon / <\mu>$ by exchanging the mean Lyapunov exponent with the largest and assuming a logarithmically weak dependence on the perturbation was dropped as inconsequential in that work.  Thus, the Lyapunov regime gives classical behavior for the quantum system.

This is a rather curious and interesting situation.  For the weak perturbation regime, the predictability/reversibility time scale extends beyond the Heisenberg break time defined by time-energy uncertainty relation and the mean level spacing.  In the medium perturbation strength range, it is still valid to an long time scale.  Clearly, the quantum dynamics remains predictable and reversible on very long time scales and is good far longer than the classical dynamics.  However, for very strong perturbations, the quantum time scale converges to the short classical scale.  Alternatively, one can imagine what appears a priori to be a very weak perturbation or uncertainty, but is fixed.  Then in the limit of $\hbar\rightarrow 0$, the quantum $\tau_r$ always recovers the classical behavior for small enough $\hbar$.  Unitary propagation may seem like it should always be stable, predictable, and reversible, but for $\hbar^2$ less than $\epsilon^2 K(E)/<\mu>$ there is a problem.  We return to this issue where discussing the paradox resolution.

Finally, there is an Ehrenfest time scale for the quantum dynamics~\cite{Berman78,Berry79}.  One way to define it here is as the latest time at which interference phenomena entering the quantum dynamics can be delayed no matter how the initial state is chosen.  Of course, interference marks a divergence between quantum and classical expectation values, hence the name Ehrenfest.  The logic and argumentation can be mapped onto that given for the classical mixing time if one equates Planck's constant $h$ with the phase space volume $4\pi \sigma_x \sigma_p$.  Thus, 
\begin{equation}
\tau_E = \frac{d}{h_{KS}}\ln \frac{{\cal V}(E)}{h} = \frac{d}{h_{KS}}\ln {\cal N}(E) = \frac{1}{\langle \mu \rangle}\ln {\cal N}(E)
\end{equation}
where here ${\cal N}(E)$ is the number of Planck cells per degree of freedom, i.e.~the effective number of quantum states per degree of freedom.  With an equivalence of the classical localization scale and Planck's constant, the classical mixing and quantum Ehrenfest time scales are identical.

\subsection{The semiclassical dynamics of wave packets and the reversibility paradox}

In principle, the ultimate semiclassical theory for wave packet propagation (excluding various kinds of uniformizations to account for coalescing saddle points) is generalized Gaussian wave packet dynamics~\cite{Huber87,Huber88}.  It is fully equivalent to time-dependent WBK theory, but it requires complexified position and momentum coordinates.  There exists a slightly less accurate semiclassical approximation that relies exclusively on real trajectories, and it allows for a much more intuitive picture~\cite{Tomsovic91b,Oconnor92,Tomsovic93}.   The set of trajectories for chaotic systems in those works are the heteroclinic orbits that connect the analogous local area of phase space of the initial Gaussian to the final local area.  They lie on the intersections of the unstable manifold of the central trajectory of the phase space surrounding the initial wave packet with the stable manifold of the central trajectory of phase space surrounding the final wave packet.  By using them, one can construct the quantum dynamics in all detail very accurately, increasingly so as $\hbar\rightarrow 0$.  The expression for cross correlation functions is~\cite{Oconnor92,Tomsovic93}
\begin{eqnarray}
\label{eq:qm}
{\cal C}_{\beta \alpha}(t) &=& \left\langle \Phi_\beta | U (t) | \Phi_\alpha\right\rangle \nonumber \\
&\approx& \sum_{\kappa} {\cal C}^\kappa_{\beta \alpha}(t) = \sum_{\kappa}\left\langle \Phi_\beta | U_\kappa(t) | \Phi_\alpha\right\rangle
\end{eqnarray}
where $\alpha$ denotes (${\bf x}_\alpha=\langle {\bf x}\rangle,{\bf p}_\alpha=\langle {\bf p} \rangle$) and likewise for $\beta$ and the final wave packet.  The index $\kappa$ labels heteroclinic orbits that begin in the phase space neighborhood of (${\bf x}_\alpha,{\bf p}_\alpha$) and end in the neighborhood of (${\bf x}_\beta,{\bf p}_\beta$) for fixed time $t$.  The unitary propagator $U_\kappa(t)$ is constructed with the classical action and Maslov index of $\kappa$ and follows by a linearization of the local dynamics using its stability matrix.  It turns out that for each heteroclinic orbit, there is a one-to-one correspondence with saddle points of the generalized Gaussian wave packet dynamics method, and a heteroclinic contribution is nearly identical to that of its associated saddle point~\cite{Pal15}.  This gives the complete connection with classically allowed transport, but tunneling and diffraction would require extensions.  Although, the use of the true saddle points would be preferable, for our purposes the heteroclinic orbit sum is more than accurate enough.

It has been shown that the correspondence time scale for which this construction works is algebraic in inverse $\hbar$.  In greater detail, the semiclassical times scales of validity are algebraic, but not universal and last at least as long as~\cite{Oconnor92,Tomsovic93,Sepulveda92}
\begin{equation}
\tau_c = \left\{\begin{array}{ll}
\frac{c_0}{\mu\hbar} & {\rm bakers\ map} \\
\frac{c_0}{\mu \hbar^{1/2}\ln \frac{c_1}{\hbar}} & {\rm stadium\ billiard} \\
\frac{c_0}{\mu \hbar^{1/3}} & {\rm kicked\ rotor} \\
\end{array}\right.
\end{equation}
These long time scales vastly exceed the predictability/reversibility time scale for the classical orbits which are being used to construct the quantum dynamics in great detail.

\subsubsection*{The semiclassical reversibility paradox}

Therefore, for strongly chaotic systems, classical dynamics is unstable, and unpredictable/irreversible after a logarithmically short time scale, whereas quantum dynamics is stable,  and predictable/reversible to at least a long, linear time scale (with the exception of very strong perturbations noted above).  Furthermore, semiclassical dynamics, using only classical trajectory information, reconstructs quantum dynamics more and more accurately in the limit of $\hbar \rightarrow 0$ for at least algebraic times scales far beyond the time that the classical dynamics has become unpredictable and irreversible.  How can that be?

\section{Illustration with the kicked rotor}

The kicked rotor has long been an extraordinary paradigm for studies of chaotic systems.  From a classical perspective, it has many desirable properties.  Its diffusion constants, action diffusion constants, and Lyapunov exponents can be calculated analytically~\cite{Chirikov79,Lakshminarayan99,Tomsovic07}.   Quantum mechanically, the kicked rotor's quantization is straightforward and well adapted for efficient numerical studies.  It is quite useful as an illustration of the above considerations.

\subsection{The classical dynamics}

A general kicked rotor is a mechanical-type particle constrained to
move on a ring that is kicked instantaneously every multiple of a unit
time, $t=j\tau$.  Supposing the radius of the ring to be $1/2\pi$ and
\begin{figure}
\begin{center}
\includegraphics[height=5 cm]{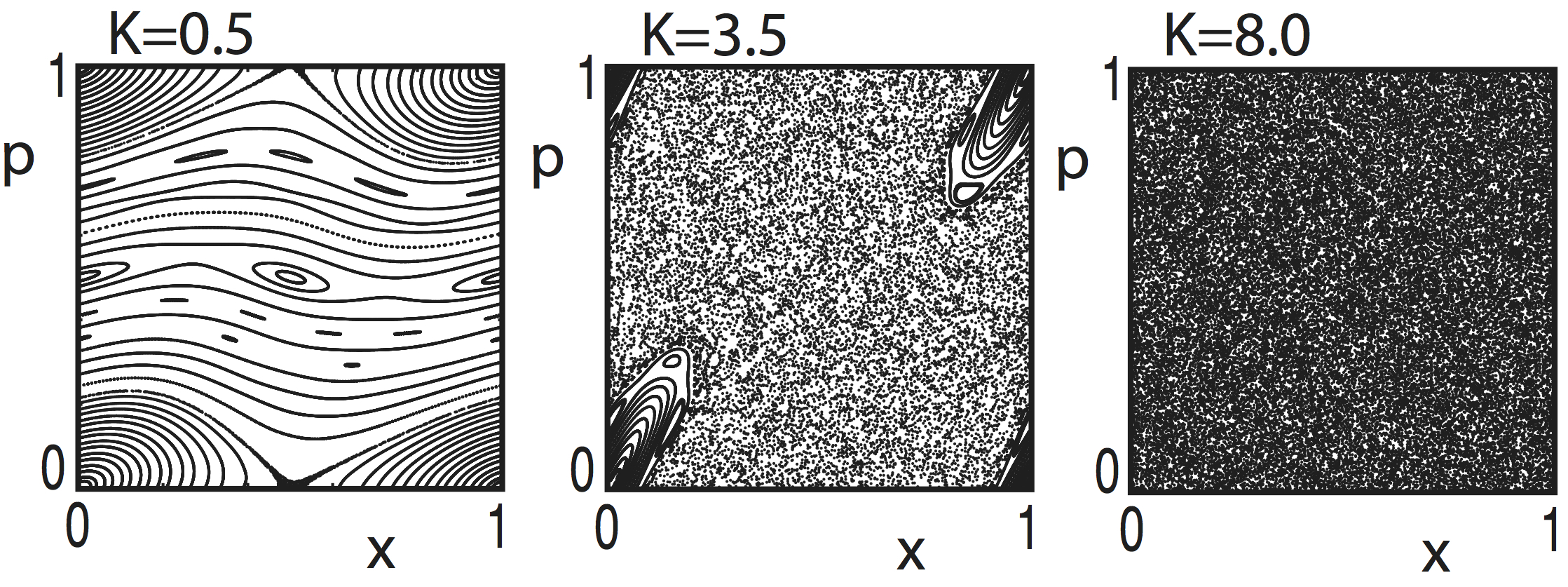}
\end{center}
\caption{Three Poincar\'e surfaces of section obtained for different values of the coupling $K$ by iterating the map Eqs.~(\ref{kreq}) for one ($K=8.0$) or a few ($K=0.5$ and $K=3.5$) initial conditions.  The left section illustrates nearly integrable dynamics.  The middle section illustrates mixed phase space dynamics with only one significant remaining region of regular motion.  The right section illustrates chaotic dynamics.  For values of $K>> 5.0$, all the surfaces of section have the same global appearance.  One can find very tiny regions of regular motion embedded in the chaos with good search methods, but they represent a microscopic proportion of the total phase space volume.} 
\label{fig:sos}
\end{figure}
$\tau=1$, the Hamiltonian takes the form 
\begin{equation}
\label{krg}
H(x,p) = \frac{p^2}{2} -\dfrac{K}{ 4\pi^2 }\cos \left(2\pi  x \right)  \sum_{j=-\infty}^\infty \delta(t-j)
\end{equation}
From $H(x,p)$, mapping equations relating the position and momentum
$(x_{j+1},p_{j+1})$ of the particle just before the $(j+1)$'th kick to
the one $(x_{j},p_{j})$ just before the $j$'th kick are obtained as
\begin{eqnarray}
\label{kreq}
p_{j+1} &=&  p_j -\dfrac{K}{ 2\pi}\sin \left(2\pi  x_j \right)  \nonumber \\
x_{j+1} &=& x_j + p_{j+1}  \; .
\end{eqnarray}
Figure \ref{fig:sos} shows the transition from integrable (regular)
dynamics to chaotic dynamics as $K$ increases.  Beyond
$K\approx 5$ the standard map is considered to be largely chaotic,
although it is also not proven to be completely chaotic for any value
of $K$.  The Lyapunov exponent of the map is given accurately by $\ln(K/2)-(K^2-4)^{-1}$~\cite{Tomsovic07}.

\begin{figure} 
\begin{center} 
\includegraphics[height=8 cm]{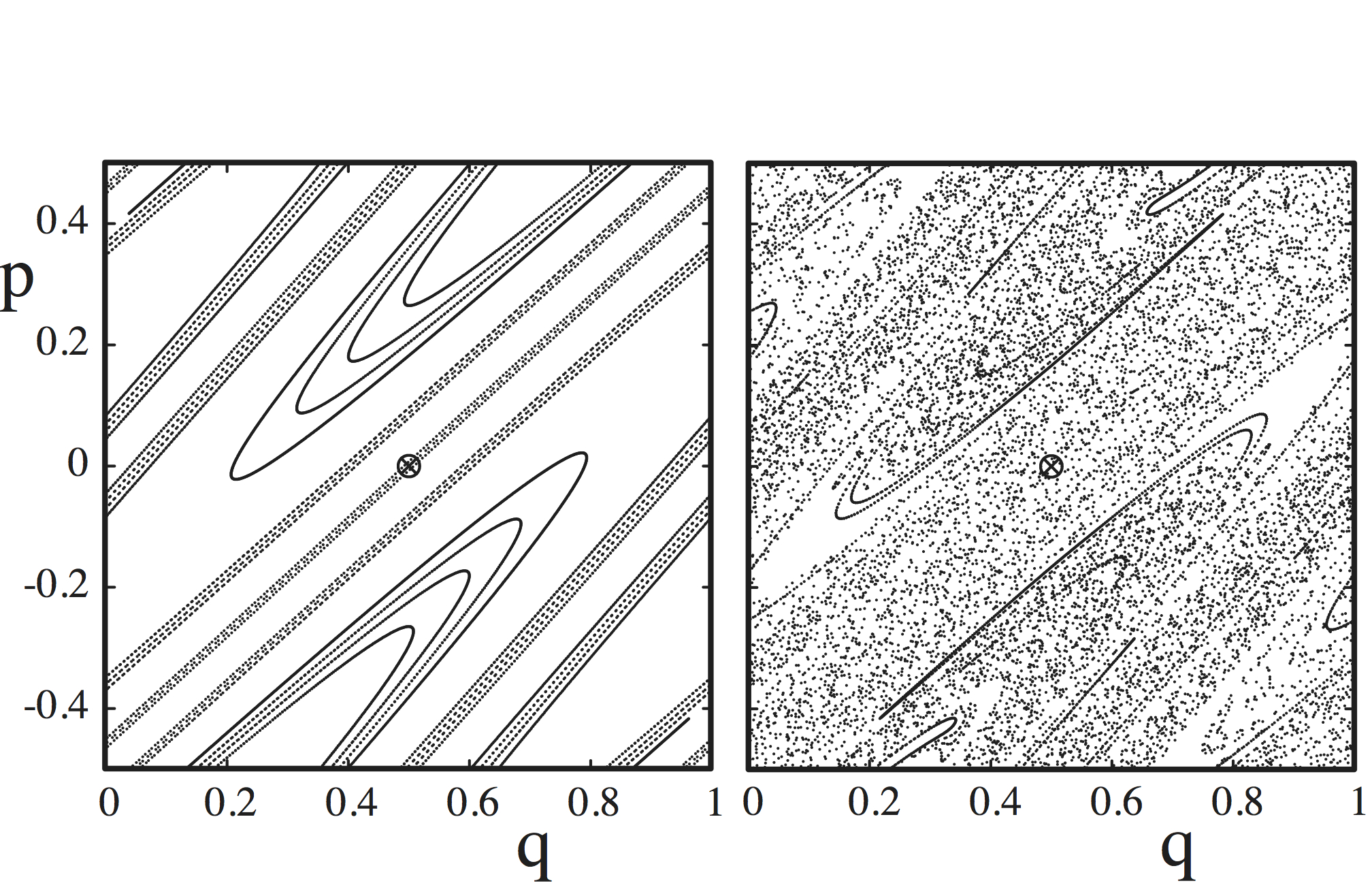} 
\end{center}
\caption{Illustration of classical mixing.  For $K=10$, the small circular volume ($1/1000$ of the total) is propagated forward in time $t=3 , 5$ on the left and right respectively.  From Eq.~(\ref{mixing}), the mixing time is approximated as $\tau_m = \ln 1000 /\ln 5 = 4.3$, which is consistent with the results shown.}
\label{dynamics}
\end{figure}

The first property to illustrate is mixing.  A small volume is propagated forward for some time to show how fine a scale it covers the available phase space.  This is drawn in Fig.~\ref{dynamics}.
One sees that the small volume spreads rapidly to finer and finer scales with increasing time.  The approximate mixing time for this case $\tau_m=4.3$ suggests that most of the phase space cells of size $1/1000$ of the square should be intersected on the right frame, but not yet on the left, which is roughly correct.

\begin{figure} 
\begin{center} 
\includegraphics[scale=.75]{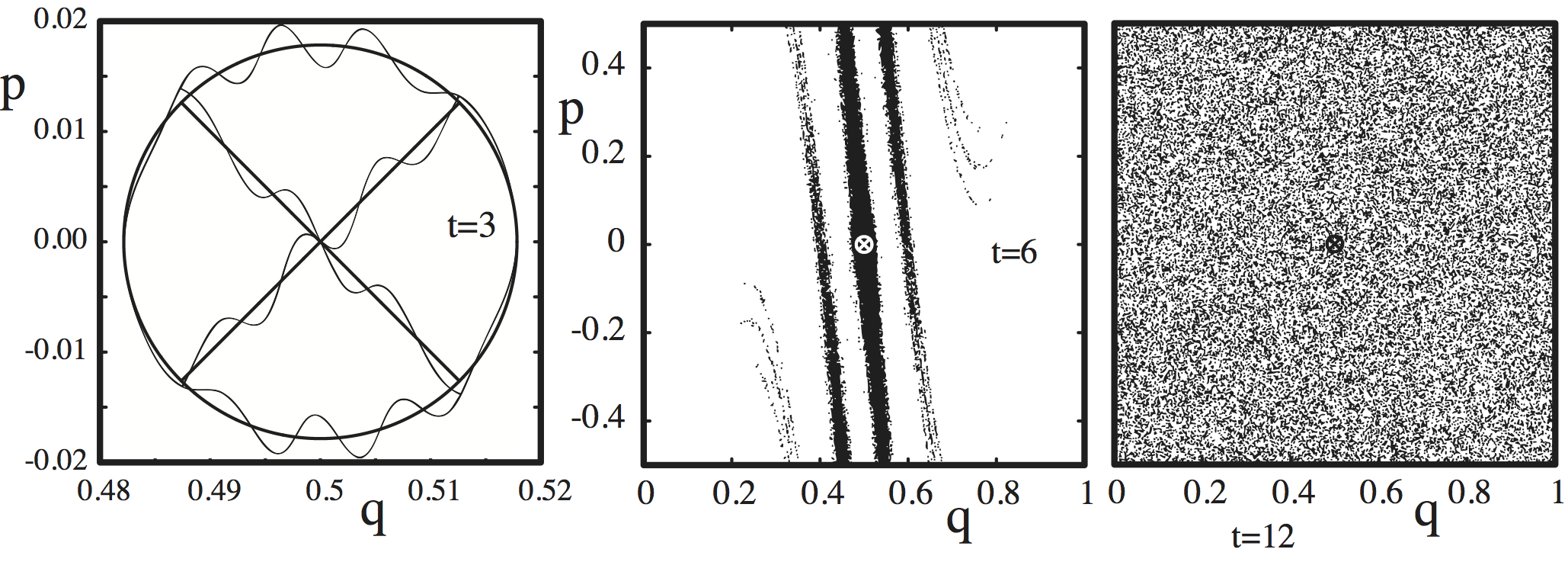} 
\end{center} 
\caption{Reversing classical dynamics.  A circular density of small volume ($1/1000$) centered at $(x,p)=(0.5,0.0)$ is propagated forward in time with a kicking strength $K=10$ and then reversed for the same amount of time with $K=10.0001$.  From left to right, the forward propagating times are $t=3,6,12$ respectively.  The reversal time scale from Eq.~(\ref{reversal}), with $\alpha=1$ is $\tau_r=\tau_p = 7.2$.  The phase space for $t=3$ is magnified to show more clearly the initial localized region.}
\label{reverse}
\end{figure}

The classical property of reversibility and predictability are illustrated by propagating with two different kicking strengths separated by $\epsilon = \delta K/K=10^{-5}$ as shown in Fig.~\ref{reverse}.  It is most visually striking to 
consider forward and reverse propagation with $H_0$ and $H_\epsilon$ respectively, and make the comparison to the initial density.  One sees in a blow-up of the phase space near the initial conditions that at $t=3$, the localized density is nearly reconstructed.  At $t=6$, there is still some ``memory'' left of beginning localized, but by $t=12$, the density is ergodically covering the full phase space and lacks any sign of having begun as a small localized density.  The time scale estimate for reversibility, $\tau_r=7.2$, is consistent with these results.

\subsection{The quantum dynamics}

\begin{figure}
\includegraphics[scale=.7]{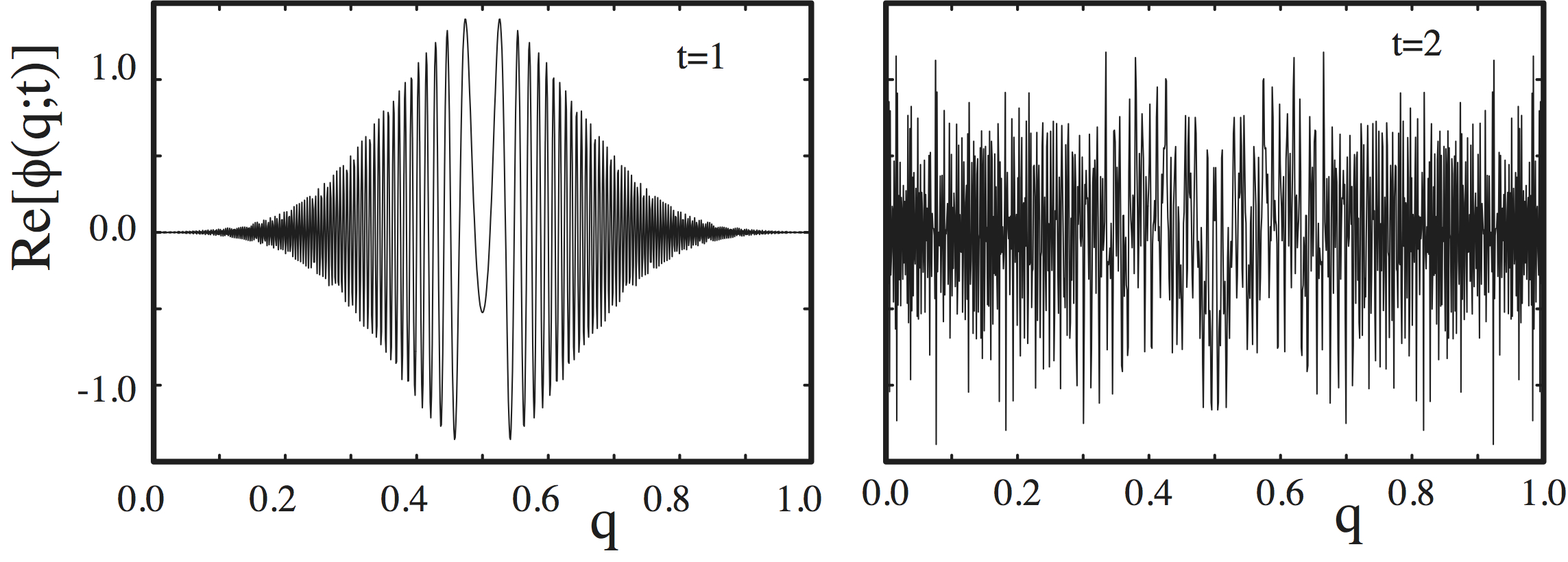} 
\caption{Forward propagation of quantum dynamics.  A wave packet centered at $(q_0=0.5,p_0=0.0)$ is propagated forward in time with a kicking strength $K=10$ and $h^{-1}=N=1000$.  For this wave packet, $t=2$ indicates the beginning of interference becoming important.  Although, the Ehrenfest time is $\tau_E=4.3$, the initial wave packet is not optimized for delaying interference to the latest possible propagation time.  Hence, nonclassical features commence before $\tau_E$.}
\label{quantumf}
\end{figure}

The quantized version of the standard map relies on the single time step propagator $\hat U$ which relates the quantum wave function of the rotor $\Phi(x;t)$ just before the $t$'th kick to the one $\Phi(x;t+1)$ just before the $(t+1)$'th kick
\begin{equation}
\Phi \left(x ; t+1 \right) =  \hat U \Phi \left(x; t\right) \; .
\end{equation}
The position variable can take on only quantized values and forms a complete basis for the quantization.  In 
this discrete basis the propagator is an $N \times N$ matrix given by 
\begin{equation}
\label{quantummap}
       \langle n |\hat U| n^{\prime} \rangle \, =\, \frac{1}{\sqrt{i N}}
\exp\left[\frac{i \pi}{N} (n-n^{\prime})^2\right]
\exp \left(i \frac{K N}{2 \pi} \cos\left[\frac{2 \pi}{N} (n+a)\right] \right).
\end{equation}
where $n,n^{\prime}\, =\, 0, \ldots, N-1$ are integers labeling the allowed discrete positions.  The parameter  $K$ is the same kicking strength as for the classical rotor, Planck's constant is given by $h=N^{-1}$, and the parameter $a$ is a purely quantum phase (associated with the choice of boundary conditions).  Taking $a=1/2$ leads to maximal quantum symmetry.  Propagation of any initial state follows by repeated multiplication by $\hat U$ to the time desired.  If the initial state is expressed in a position basis, it is given as a column vector and Eq.~(\ref{quantummap}) gives the form of $\hat U$'s matrix elements to be used in the multiplication.  Figure~\ref{quantumf} shows the first two forward time steps of 
the wave packet that corresponds to the phase space circle of initial conditions used for illustration in the previous section.  Already by $t=2$, the wave packet is fully delocalized in the position representation and beyond its correspondence time with classical expectation values.  In this particular case, the interference shows up at half the Ehrenfest time because the initial state was not optimized to delay interference phenomena to the latest possible time.

\begin{figure}
\begin{center} 
\includegraphics[height=8 cm]{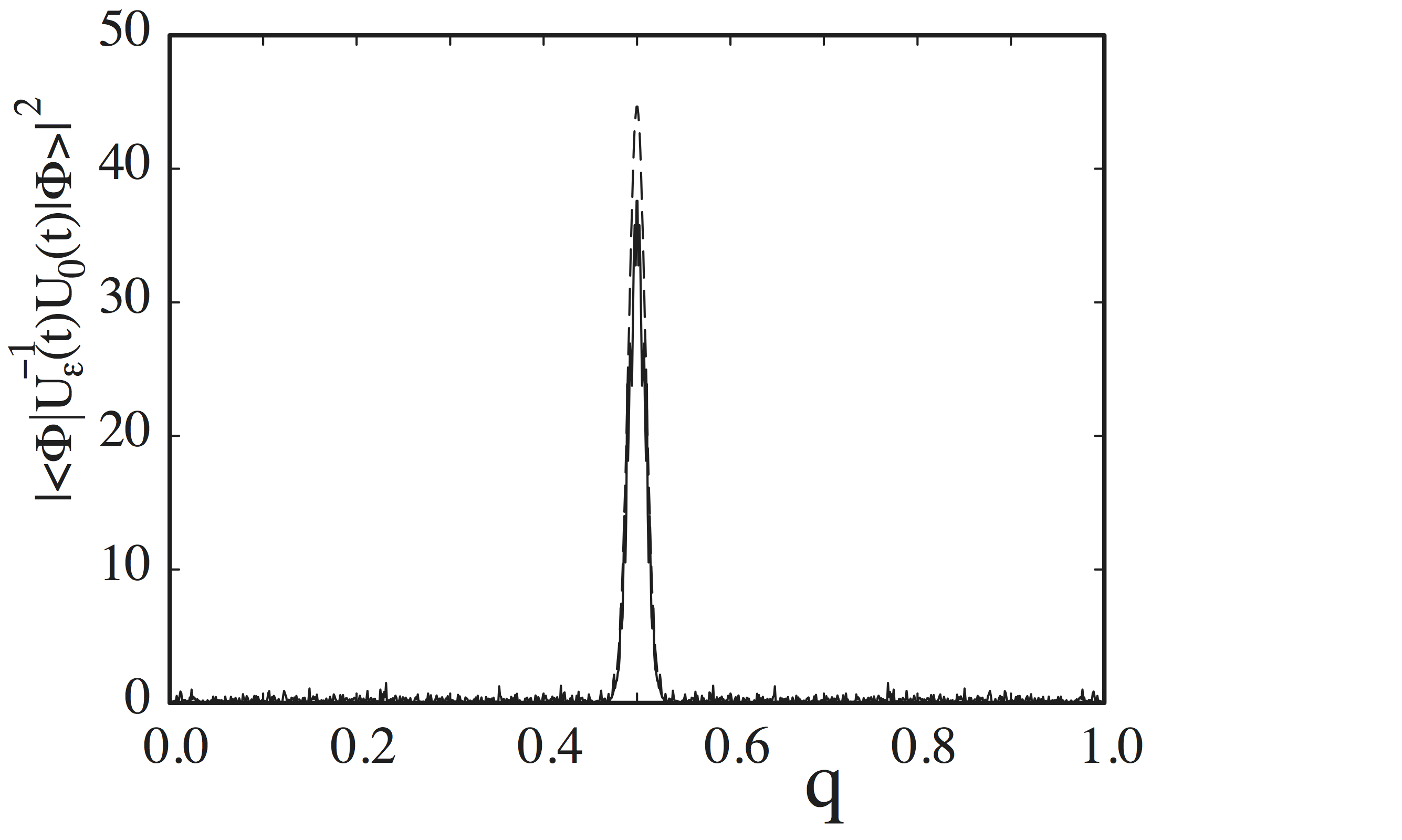} 
\end{center} 
\caption{Forward and reversed quantum propagation.  An initial wave packet centered at $(x_0=0.5,p_0=0.0)$ is propagated forward in time with a kicking strength $K=10$ and $h^{-1}=N=1000$ for $500$ time steps.  It is then reverse propagated for $500$ time steps with $K=10.001$ ($\epsilon=10^{-4}$).  The absolute square of the initial (dashed line) and forward-reversed (solid line) wave packets are shown.  The absolute square of their overlap is $0.77$.  In this example, even for times of the order of a hundred times the classical reversibility time scale, the quantum propagation is nearly reversible.}
\label{revquantum}
\end{figure}

Next consider the forward and reversed quantum propagation for this same wave packet.  Analogous to Eq.~(\ref{classpredict}), the quantum dynamical quantity of interest is
\begin{equation}
F(t)  = \left| \left\langle \Phi_\epsilon(t) | \Phi _0 (t)\right\rangle \right|^2 = \left| \left\langle \Phi | U^\dagger_\epsilon(t) U(t) | \Phi \right\rangle \right|^2
\label{fidel}
\end{equation}
which is a measure of how similar the propagation forward in time of a quantum wave packet with slightly different Hamiltonians (or forward-reversed propagation of a wave packet, no difference).  Figure \ref{revquantum} shows that even for $t=500$, the wave packet relocalizes better than the classical density managed only for $t=5$.  The results are in quantitative agreement with semiclassical theory using the functional forms given in~\cite{Cerruti03} and the times scales of Eq.~(\ref{tscales})
with the analytic form of the classical action diffusion constant~\cite{Lakshminarayan99}
\begin{equation}
K(E)\approx\frac{1+2J_2(K)}{4(2\pi)^4}
\end{equation}
$g=4$, and $\tau_H=N$ for a Floquet system.  For every further decrease of $\epsilon$ by a factor $10$, the time scale ratio of quantum reversibility to classical reversibility increases another factor $100$.

\section{Resolution of the paradox}

As mentioned in the beginning, all through the time period in which the classical dynamics are unpredictable and irreversible, but the quantum dynamics are stable and reversible, a semiclassical approximation of the quantum dynamics gives a very accurate reconstruction.  The resolution of the paradox becomes clear by looking in greater detail at the ingredients of the semiclassical approximation.

\subsection{Heteroclinic tangles}

\begin{figure}
\begin{center} 
\includegraphics[height=7.3 cm]{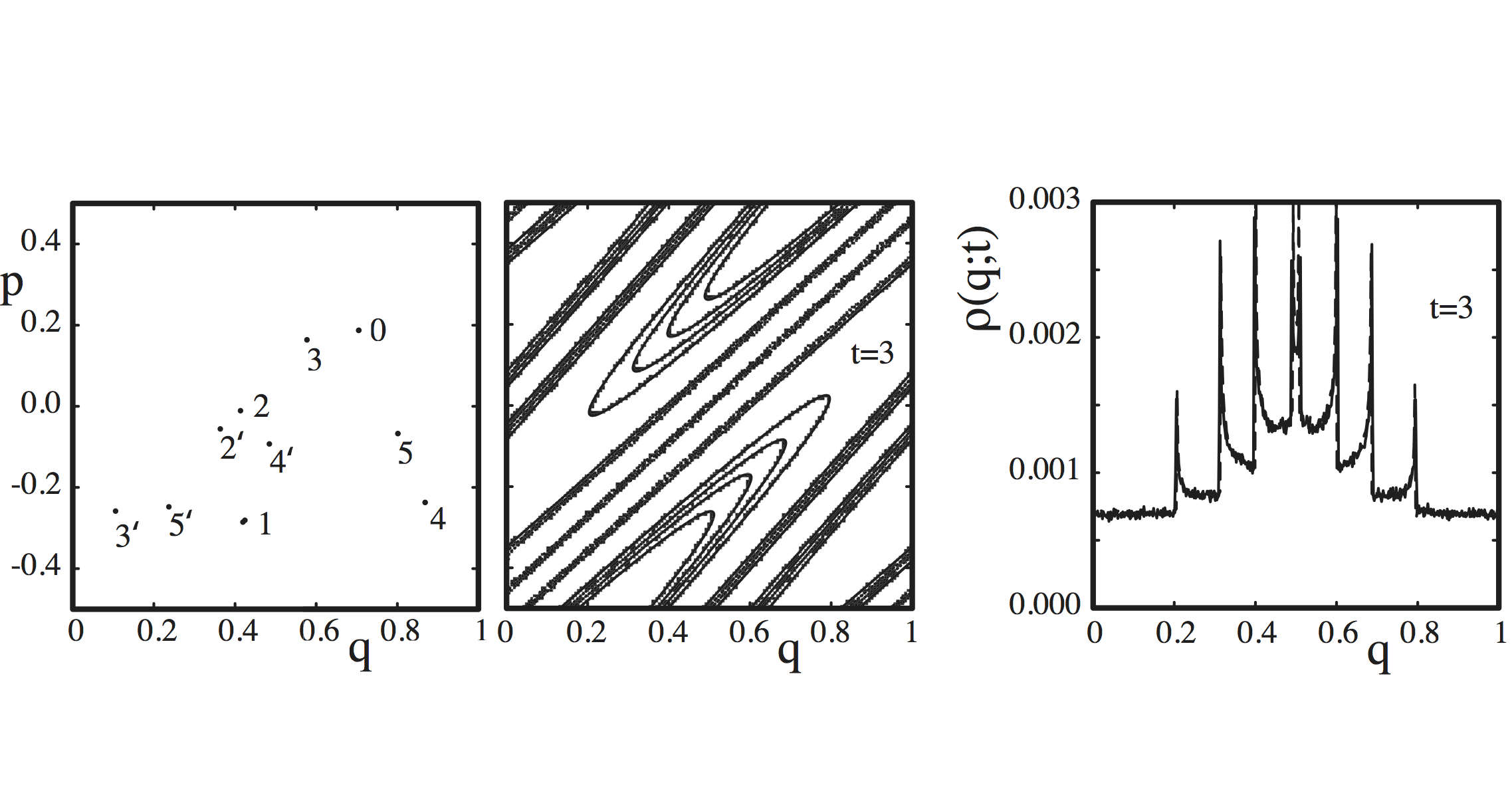} 
\end{center}
\caption{Unpredictable and predictable in classical dynamics.  Using both $K=6.00$ and $K=6.02$ in the left panel, an initial condition is propagated forward in time to $t=5$.  By $t=3$, the two resulting trajectories with one denoted by a $^\prime$ have a large separation, spanning the scale of the full phase space.  However, the unstable manifolds of the two trajectories, shown in the center panel, are almost identical and cannot be distinguished in the figure.  Their calculation is stable.  Furthermore, the unstable manifold projections into configuration space in the right panel show that there are the same number of caustics and that they have barely shifted in location.}
\label{ffproprfig}
\end{figure}

Consider a strong enough perturbation that the exponential divergence in the classical dynamics is visible at very short times.  In the left panel of Fig.~\ref{ffproprfig}, an initial condition denoted by $0$ is propagated $t=5$ iterations under the action of two kicking strengths.  Although by $t=3$, the
individual orbits are far apart and quickly become unpredictable, the collection of orbits that make up the unstable manifolds for the two kicking strengths are not distinguishable on the scale visible in the figure.  The manifolds are predictable even if one cannot calculate properly any of the orbits of which they are comprised.  There is a theorem by Moser that proves a neighborhood of convergence exists in a normal form transformation~\cite{Moser56}, which was extended to all times by da Silva Ritter et al~\cite{Silva87}.  Effectively, any errors or deviations in an orbit calculation are reduced exponentially back onto the unstable manifold forward in time.  The way that two orbits diverge exponentially from each other is almost exclusively to slide along the unstable manifold away from each other.  The unstable manifold construction is effectively impossible using the inverse dynamics, thus it is a one-way construction.  These considerations extend to perturbed systems as well.  A small perturbation is like an error or deviation that is being constantly reduced exponentially.   Thus, the unstable manifold of an orbit, say a periodic one, of two systems $\epsilon$ apart have nearly the identical unstable manifold.  A shorthand name for this strong structural stability in chaotic systems~\cite{Ott02} is manifold stability~\cite{Cerruti02}.  Similarly, the same conditions apply for the stable manifolds inversely in time.  In fact, this stability can be used to calculate heteroclinic orbits in a much more stable way than by using Hamilton's equations.  It suffices to take advantage of this structural stability and the fact that heteroclinic (or homoclinic) orbits lie on stable and unstable manifold intersections~\cite{Li15}.

An important consequence of this manifold stability follows from the association of phase space areas to classical actions~\cite{MacKay84a, MacKay87, Meiss92}.  Given that the unstable manifold is accurately constructed, then if the endpoints of a heteroclinic (homoclinic) orbit are known as typically happens in semiclassical methods, so is its classical action by measuring the area.  One would just measure the area under the unstable manifold using the initial and final points as the limits of integration.  Not knowing the history of the orbit would cause no problems.  Similar arguments can be made for counting caustics by following the manifold and for obtaining the stability parameters of the orbit.  It turns out rather ``conveniently'' that all the classical information required for the heteroclinic (homoclinic) construction of the quantum dynamics given by semiclassical theory is stably calculable even over propagation times where the specific orbits are not.  Furthermore, the number of terms in the summation at any given time, i.e.~how many heteroclinic (homoclinic) orbits contribute is also stable and a predictable quantity.  All of this relies on the well-known, strong structural stability of unstable and stable manifolds~\cite{Ott02} and it is necessary for the semiclassical construction to work.  The converse is also true, given that the semiclassical construction works, with just a little bit of reflection it would be clear that there had to exist in the classical dynamics a structural stability that would enable all the needed classical information to be stable and predictable.  If the structural stability had not been known, the validity of the semiclassical theory would have suggested its existence and the search for it.

\subsection{Classical perturbation theory}

To construct a propagating wave packet semiclassically, only the unstable manifold is needed.  Each point on the manifold is an orbit propagated for a time $t$ that contributes to the evolved quantum state $\Phi({\bf x};t)$ at the position of its endpoint ${\bf x}_t$, weighted by where it began in the wave packet $\Phi({\bf x}_0;0)$ at $t=0$ and by its stability properties.  Comparing the propagation of a perturbed and unperturbed case requires comparing every orbit on one unstable manifold with its corresponding orbit on the other perturbed one.  They are matched so that each member of such a pair has the same final position.  Their action difference is given by first order classical perturbation theory to be
\begin{equation}
\label{perteq}
\Delta S = - \epsilon \int_0^t {\rm d}\tau H_1[{\bf x}(\tau), {\bf p}(\tau)]
\end{equation}
where the path of integration is over the unperturbed orbit.  Naively, unaware of manifold stability, one might have expected this quantity to diverge exponentially fast with increasing time due to the exponential separation of orbits in chaotic systems, and for quantum dynamics of chaotic systems to be hypersensitive to perturbations.   This would have \begin{figure}
\begin{center} 
\includegraphics[height=6 cm]{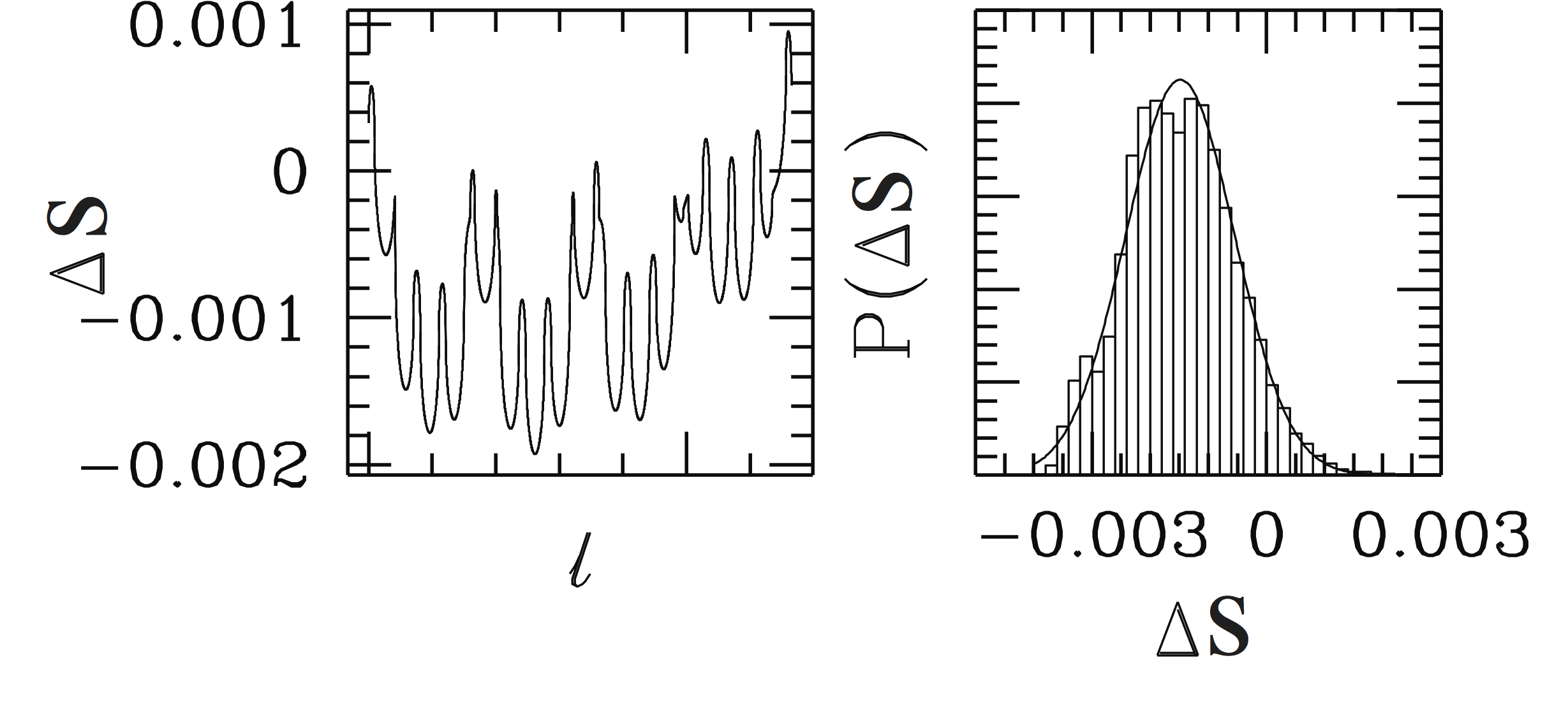} 
\end{center} 
\caption{Action changes along the unstable manifold. The solid line in the the left panel is the action difference for the two unstable manifolds as shown in Fig.~\ref{ffproprfig}, except that the manifolds are propagated $t=4$ iterations.  The dashed line in the left panel is the perturbation result, Eq.~(\ref{perteq}), which is too accurate to distinguish from the solid line.  The variable ${\it l}$ represents ``length'' along the manifold, and therefore is single valued.   The right panel is a histogram of the action changes for $t=8$ iterations along the manifold. The solid line is a Gaussian function, whose width is given by the classical action diffusion constant, Eq.~(\ref{diffuse}).  Modified from Fig.~3 of~\cite{Cerruti02}.}
\label{pertheor}
\end{figure}
been due to the action difference showing up as an exponentiated phase.  It would also seem to be limited to logarithmically short time scales.  Instead, the manifold stability is strong enough to eliminate the exponential sensitivity completely, and, in fact, the first order classical perturbation term is valid on algebraically long time scales.  It is responsible, after further argumentation, for introducing the classical action diffusion constant.  There has been discussion of hypersensitivity to perturbations~\cite{Schack93,Schack96} for quantum chaotic systems, but the context is different than that being discussed here.  

A nice illustration of the accuracy of perturbation theory is given in~\cite{Cerruti02}; a modified version of their Fig.~3 is shown as Fig.~\ref{pertheor}.  It is possible to measure the ``distance'' along the kicked rotor's unstable manifold, $\it l$, and use it to order the pairs of trajectories mentioned above.  In the left panel of Fig.~\ref{pertheor}, the actual action difference and the first order perturbation theory approximation are plotted versus $\it l$.  The two curves are so similar that one cannot distinguish them.  In the left panel, it appears as though only a single curve has been drawn.  This verifies the excellent accuracy of classical perturbation theory for a strongly chaotic system.  There are some interesting properties of the action difference curves.  Given the relation between action differences and phase space areas, each local maximum or minimum in the curve occurs at a crossing point of the perturbed and unperturbed unstable manifold.  It is at those locations where the sign changes from a positive action difference to negative or vice versa.  In fact, the difference of the value between a minimum and the next maximum, or maximum and the next minimum, is the area contained between the two manifolds from the loop formed by successive crossing points.  We'll return to these areas just ahead, here let's call the set $\{\delta{\cal S}_j\}$.  Following the behavior of the perturbed and unperturbed manifolds over a great distance leads to a statistical behavior in their action differences.  This is shown in the right panel of Fig.~\ref{pertheor}.  There is a classical action diffusion in action differences, and hence a central limit theorem for their distribution; i.e.~a Gaussian limiting density with a variance given by the diffusion constant, Eq.~(\ref{diffuse}).

\subsection{The overlap integral}

The fidelity, Eq.~(\ref{fidel}), can be written as an overlap integral over position space by inserting a complete set of position vectors
\begin{equation}
F(t) = \left| \int_{-\infty}^\infty{\rm d}{\bf x}\ \Phi_\epsilon^*({\bf x};t) \Phi_0({\bf x};t)   \right|^2
\end{equation}
The semiclassical evaluation of this overlap integral depends on the action differences displayed in Fig.~\ref{pertheor}.  As noted in the previous section, the manifold stability allows for very small, stable action differences.  For a sufficiently small value of $\epsilon$, they are below the scale of Planck's constant.  More precisely, for values of $\epsilon$ for which $ \gamma^2 > \frac{d}{h_{KS}\cdot \tau_H}$ [see Eqs.~(\ref{gamma},\ref{tscales})], the overlap cannot be evaluated by the method of steepest descents approximation.  In this limit, the areas $\{\delta{\cal S}_j\}$ are so small that the conditions for a saddle point analysis are continuously violated moving along the manifold.  However, given it's Gaussian fluctuation density, the integral is straightforward to evaluate, and the only information surviving the overlap integral is the aforementioned diffusion constant, Eq.~(\ref{diffuse}).  In this regime, the quantum and classical predictability and reversibility properties are extremely different, and nevertheless, for the quantum reversibility properties the purely classical information in the form of the classical action diffusion constant determines everything.

If this overlap integral were amenable to a saddle point analysis, then the reversibility properties of the quantum and classical fidelity would have to be identical, because one could follow the orbits forward and backward in time using the saddle point approximation the entire way.  In fact, the information about this issue is exactly contained in the criteria on $\gamma$ just mentioned.  For example if $\epsilon$ is fixed, in the limit of $\hbar\rightarrow 0$ there comes a point where the typical or majority values of the $\{\delta{\cal S}_j\}$ become greater than the value of $\hbar$.  The crossover occurs just where the relation above switches to $ \gamma^2 < \frac{d}{h_{KS}\cdot \tau_H}$.  That is the  point at which the method steepest descents becomes accurate again.  Thus, it is also the point at which the quantum reversibility properties again become identical to the classical, and the classical action diffusion constant is no longer relevant.  In other words, that is why it has to be the criterion for whether the quantum reversibility properties are in the Lyapunov regime, Eq.~(\ref{tscales}), which is the only regime where they match the classical reversibility properties.  In the final analysis, it is the structural stability of chaotic systems, which enables the predictability and reversibility of quantum and classical dynamics to differ.

\section{Discussion}

The quantum and classical dynamical worlds are very different, and some of their intimate connections are not always fully appreciated.  That semiclassical methods relying exclusively on classical information can be used to construct more and more accurate quantum propagation over algebraic time scales in the $\hbar\rightarrow 0$ limit has additional implications that do not derive from the idea that classical dynamics must emerge from quantum dynamics in the same limit.  In fact, classical dynamics emerges only logarithmically slowly from quantum dynamics.  This is not an equivalent two-way street in that it is easier to bootstrap quantum dynamics from classical than classical dynamics from quantum.  

The term paradox is used here because although it may seem odd to have the situation that information from an unpredictable dynamics fully reconstructs a predictable one, there is no contradiction.  The quantum dynamics has less information in it and interference is responsible for much of its destruction.  This is reflected in entropies.  Given that the orbits themselves are unpredictable, one might expect that the semiclassical term associated with a particular orbit cannot be so important.  However, all the terms in a semiclassical sum must be there and be accurate if the final summation is to give the correct results.  It is not a case of random numbers that can be anything so long as their sum gives the correct value.  Imagine opening a system that allows some transport pathways to escape or that modifies some paths, but not others.  The semiclassical sum has to be able to handle all of these modifications, and can only do it if each term has a meaning and value in spite of having access only to the sum for a particular situation.  Rather amusingly, the structural stability of chaotic systems takes care of this by making the needed orbit's classical information stable and predictable even though the orbit itself is not.  The number of orbits in the summation, their actions, Maslov indices, stabilities are all stable.

With respect to the reversibility properties of quantum and classical dynamics of simple chaotic systems, the inherent, strong, classical structural stability enables the reversibility properties to be far different in many regimes despite the fact that classical information predicts the quantum dynamics all throughout these regimes.  Conversely, one could deduce that given that semiclassical theories work and the reversibility properties are extremely different, chaotic systems must be structurally stable.  All other quantities being fixed, in the limit $\hbar\rightarrow 0$ there always comes a point beyond which the quantum and classical reversibility properties match.  This can also be thought of as the large perturbation regime.  In the opposite limit of a fixed $\hbar$, shrinking the perturbation always leads to a point beyond which the reversibility properties get arbitrarily different in time scales, tending to an infinite ratio of quantum to classical reversibility time scales.  The crossover points in these limits are determined by the validity of the saddle point approximation applied to the overlap integral.  It in turn depends on how similar the unstable manifolds for a system and its perturbation are.  The end result is a single universal scaling parameter, here denoted as $\gamma$, that controls the relationship between the quantum and classical reversibility behaviors.  It is completely determined by the perturbation strength, classical action diffusion constant, Planck's constant, the Heisenberg time scale, and a symmetry index.

\section*{Acknowledgments}

The author gratefully acknowledges helpful discussions with and a critical reading of the manuscript by Arul Lakshminarayan.

\bibliography{quantumchaos,general_ref,rmtmodify,classicalchaos}

\end{document}